**Dynamic defect correlations dominate activated electronic transport in SrTiO$_3$**

*Paul C. Snijders*,[1,2]* *Cengiz Şen*,[3] *Michael P. McConnell*,[2] *Ying-Zhong Ma*,[4] *Andrew F. May*,[1] *Andreas Herklotz*,[1] *Anthony T. Wong*,[1] and *T. Zac Ward*[1,]*

[1] Materials Science and Technology Division, Oak Ridge National Laboratory, Oak Ridge, Tennessee 37831, USA
[2] Department of Physics and Astronomy, University of Tennessee Knoxville, Tennessee 37996, USA
[3] Department of Physics, Lamar University, Beaumont, Texas 77710, USA
[4] Chemical Sciences Division, Oak Ridge National Laboratory, Oak Ridge, Tennessee 37831, USA

* E-mail: snijderspc@ornl.gov, wardtz@ornl.gov

Strontium titanate (SrTiO$_3$, STO) is a critically important material for the study of emergent electronic phases in complex oxides, as well as for the development of applications based on their heterostructures. Despite the large body of knowledge on STO, there are still many uncertainties regarding the role of defects in the properties of STO, including their influence on ferroelectricity in bulk STO and ferromagnetism in STO-based heterostructures. We present a detailed analysis of the decay of persistent photoconductivity in STO single crystals with defect concentrations that are relatively low but significantly affect their electronic properties. The results show that photo-activated electron transport cannot be described by a superposition of the properties due to independent point defects as current models suggest but is, instead, governed by defect complexes that interact through dynamic correlations. These results emphasize the importance of defect correlations for activated electronic transport properties of semiconducting and insulating perovskite oxides.

**Introduction**

In modern materials design, the perfection of material synthesis capabilities has led to numerous discoveries and technologies ranging from quantum Hall effects in high mobility semiconductor heterostructures[1,2] to tunable microwave dielectrics in complex oxides.[3] In these examples, point defects and impurities in the crystalline lattice are often deleterious, because they profoundly affect the electronic properties. However, precisely because of their effect on materials' properties, defects and impurities offer an extremely versatile way to significantly expand the range of phenomena exhibited by a parent material that can be used to control new properties. STO is an excellent example of this: oxygen vacancies can control resistive switching[4,5] and lead to magnetism[6] or even superconductivity;[7] strontium vacancies can induce ferroelectricity;[8] and the presence of both cation,[9,10] and anion[11,12] point defects has been used to explain emergent electronic properties near STO - perovskites interfaces. Furthermore, in complex oxides beyond STO, oxygen vacancies and other point defects can produce active sites with a high catalytic activity for e.g. oxygen reduction or evolution reactions.[13]

It is often challenging to ascertain the type of defects present, in particular when defect concentrations are low, and most point defects need to be monitored by indirect methods.[14] However, even at concentrations below the detection limit, defects can have a tremendous effect in the properties of functional materials as evidenced by the appearance of persistent photoconductivity due to low concentrations of defects.[15,16,17] Persistent photoconductivity can thus be exploited to probe low concentrations of defects, in particular those that are relevant for the electronic properties of the material. Specifically, exposing semiconductors or insulators to light with energies near and above the band gap photoexcites charge carriers, which are subsequently trapped in defect states within the gap.[15,16,17] Persistent electronic conduction happens through hopping mechanisms between these defect states, and its decay with time offers insight into the energy barriers that are involved in de-trapping the carriers responsible for the observed conductivity.[15,16,17] The altered electronic properties of defected crystalline materials are usually considered in terms of isolated, independent point defects such as the oxygen vacancies in STO mentioned above.[18] However, interactions between isolated defects can create defect complexes that have a lower formation energy than isolated point defects[19,20,21,22] and thus dominate the properties of defected materials. The static electronic structure of these composite defect structures cannot be described by a superposition of individual defect electronic spectra.[23] Moreover, the nature of the interactions also allows for dynamic correlations in response to changes in electron occupation of defect levels. These dynamic correlations should then be apparent in correlated electronic relaxations upon creating non-equilibrium electron energy distributions ubiquitous in electronic transport experiments. Such dynamic correlations are conceptually comparable to dynamic correlations that constrain structural relaxations in solids, leading to cascading relaxation processes,[24] but their influence on effective activation energies has not been recognized yet in basic electronic transport experiments.

Here we report an experimental study demonstrating that in insulating STO single crystals with low defect concentrations, photo-induced electronic transport does not follow relaxation pathways governed by a superposition of activation energies associated with independent point defects, but instead proceeds through electronic states of defect complexes that interact through dynamic correlations. We chose to study STO because it is a prototypical insulating perovskite transition metal oxide that is widely used as a substrate to grow complex oxide thin films and heterostructures. As such STO, its defects and their properties are a crucial part of the design of functional complex oxide materials and devices as borne out by the intense focus on STO since the seminal work by Ohtomo and Hwang[25] describing the high-mobility two-dimensional electron gas at STO and LaAlO$_3$ interfaces.

In our experiments, annealing treatments were designed to produce low concentrations of different types of defects. In particular, the different annealing conditions aimed to affect the chemical potential of the elements in the samples, which, in turn, is expected to result in different defect formation energies that determine equilibrium defect concentrations.[18,19,21] The decay of the photoconductivity for these samples varies from a few minutes to a few hours. We analyze fits of the decay profiles to double or triple, as well as stretched exponential (Kohlrausch) decay functions. For each type of fit, we extract the apparent activation energies. A disparity between the activation energies extracted from the two types of fits suggests that the persistent conductivity originates in clusters of interacting defects, or defect complexes, rather than in isolated and independent point defects, and that its decay proceeds subject to dynamical correlations in these defect complexes. Our results indicate that designing STO-based systems requires proper control over sample illumination, and a thorough treatment of defects, their correlations, and the resulting effective activation energies, in order to properly describe their electronic properties.

**Results and Discussion**

**Figure 1** presents the sheet resistance measured at room temperature as a function of time and illumination on the differently treated STO crystals (for treatment conditions, see **Table 1** and Section 4, Experimental Methods) plotted using the same sheet resistance scale to emphasize the different behaviors. In addition to vacancy disorder due to stoichiometric Schottky defects in STO,[26] these treatments aimed at producing low concentrations of oxygen and titanium vacancies ($V_O$ and $V_{Ti}$) and $Sr_{Ti}$ antisite defects,[21] as well as extrinsic Si impurities. The sheet resistances before photoexcitation are consistently 18 to 20 GΩ/□ for 0.5 mm thick samples. Isolated oxygen vacancies, produced by the vacuum annealing, are known to rapidly increase

*n*-type conductivity[27] due to the formation of a donor state located close to the conduction band minimum of STO.[18,28] The high resistance of our annealed samples therefore suggests that the oxygen vacancy concentration in our samples is not very large. This is corroborated by X-ray diffraction data (supplementary information) that does not show evidence of a defect-induced unit cell volume increase.

Upon illumination with below band gap light, all samples annealed in a $SiO_2$ ampoule[17] (samples A, B, C) show conventional (transient) photoconductivity as well as persistent photoconductivity that continues long after illumination has ceased (Fig. 1); only the sample subjected to annealing in an $Al_2O_3$ boat exhibits no photoconductivity upon illumination (sample D, Figure 1(d)). This suggests that the measured (persistent) photoconductivity is dependent on the environmental conditions during annealing. It is known that annealing quartz at 1200 °C in vacuum produces SiO vapor,[29] and hence it is probable that silicon (oxide) is deposited on the STO samples when annealed in $SiO_2$ ampoules. Indeed, X-ray Photoelectron Spectroscopy (XPS) detects the presence of Si on the surface of sample B, as shown in the inset of **Figure 2**. Comparing the integrated intensity of the Si 2*p* photoelectron peak to that of the Ti 2*p* peak and accounting for the order of magnitude different photoionization cross sections for the respective kinetic energies of the photoelectrons from these core levels,[30] as well as their respective inelastic mean free paths (assuming a homogeneous sample) gives an estimate of the surface Si/Ti atomic ratio of approximately 0.1. While it is not possible to reliably extract the concentration of Si that is incorporated in the STO lattice because most of the silicon observed in the XPS data could be located on the surface, silicon will diffuse into the STO crystal during a high temperature anneal.[31] These results suggest that Si defects such as Si interstitials or $Si_{Ti}$ defects[14] play a role in the observed photoconductivity upon illumination with below bandgap light.

Apart from the likely Si impurities, the presence of SrO in the quartz ampoules[17] is expected to affect the type of defects present in the STO crystals after annealing. The presence of bulk SrO with the STO crystal during vacuum annealing has been predicted to produce oxygen vacancies and $Sr_{Ti}$ antisite defects,[21] and could decrease the relatively high formation energy of Ti vacancies under reducing conditions.[32] A rudimentary estimate of Sr/Ti ratio we deduce from the Sr 3*d* and Ti 2*p* core level XPS spectra (Figure 2) is Sr/Ti = 1.5 (after correcting for the difference in photoionization cross section and inelastic mean free path for the Sr 3*d* and Ti 2*p* core levels, assuming a homogeneous sample). This suggests that excess Sr is present on the surface, and is consistent with the smaller Sr/Ti intensity ratio measured on un-treated STO (not shown). Since this off-stoichiometry is much larger than expected for STO, these data indicate that SrO has likely migrated through the gas phase onto the STO surface.

In the following we analyze in more detail the persistent photoconductivity of the STO samples that were annealed under various conditions in quartz ampoules (Figure. 1 (a,b,c)) with the aim of extracting the energies of the in-gap defect states governing the persistent photoconductivity. **Figure 3** presents the decay profiles of the photoconductivity of these crystals after the illumination was stopped. Specifically, Figure 3(a) shows the photoconductivity decay of sample A, annealed without SrO, and measured in vacuum (see also Figure 1(a)). Figure 3(b) shows the decay profile measured on sample B annealed in the presence of SrO, and measured in vacuum (see also Figure 1(b)) as well as in an oxygen atmosphere (see below). Finally, the data in Figure 3(c) were measured in vacuum on sample C that was quench-cooled from high temperature in the presence of SrO (see also Fig. 1(c)), which should result in an increased defect concentration as compared to an equilibrium cool-down. Clearly the decay rates of the photoconductivity in these samples differ significantly. The longest persistent photoconductivity is observed in samples that were annealed in the presence of SrO, suggesting that, apart from possible Si impurities, intrinsic defects such as Sr$_{Ti}$ antisite defects or Ti vacancies[21,32] could be essential for long-term persistent photoconductivity in STO. Although in the quenched sample C (Figure 3(c)) the concentration of these defects is expected to be even larger as a higher temperature equilibrium concentration is frozen into this sample, the persistence of the photoconductivity is considerably shorter in the quench-cooled sample (Figure 3(c)) than in the slowly cooled samples (Figure 3(b)). This unexpected finding suggests that an increased defect concentration leads to changes in the nature of the effective trapping states.

In Figure 3(b) a second data set is also plotted, which was illuminated and measured in an oxygen atmosphere to assess the influence of oxygen vacancies. Apart from the vacuum annealing treatment which produces a low concentration of oxygen vacancies, exposure to intense synchrotron UV light in vacuum is known to produce a high concentration of oxygen vacancies near the STO surface, creating a 2-dimensional electron gas (2DEG).[33,34] Exposing such oxygen-deficient STO samples to oxygen fills the oxygen vacancies again.[35] It is unlikely that persistent photoconductivity in our experiments is caused by such a 2DEG effect because of the much lower intensity and energy of the light used in this work (3.06 eV in this work vs. ~50 eV in Refs. 33,34). This is corroborated by the data in Figure 3(b) which shows that the presence of oxygen does not destroy persistent photoconductivity. The reduction in the persistence by approximately a factor of 2 shows that oxygen vacancies are likely to play a role, however, they are evidently neither a dominant nor independent cause of the persistent photoconductivity as the photoconductivity still persists for orders of magnitude longer than the conventional transient photoconductivity. Note that if the effect of the oxygen atmosphere would be limited to filling (independent) oxygen vacancies in only a surface layer of a sample that is oxygen deficient throughout the bulk, the measured resistance should increase proportional to the decreased volume of oxygen-deficient and semiconducting STO. This is not observed, indicating that the persistent conductivity occurs throughout the bulk of the sample consistent with the findings in Ref. 17. Moreover, in a system with independent defects, the characteristic relaxation timescale should not change, because only the quantity of one type of defect is changing, while the types of the defects present in the sample remains the same. This is clearly inconsistent with the observed data in Figure 3(b).

To gain quantitative insight into what defects dominate the observed behavior, an experimental determination of the energies associated with the defects that dominate the persistent conductivity is necessary. To this end, we first fitted the time dependent conductivity σ(t) using a sum of independent exponential (Debye) decays,[36] i.e. using the common assumption that the data can be described using a superposition of independent defects; $\sigma(t) = \sigma_0 + \sum_{i=1}^{n} A_i e^{-\frac{t}{\tau_i}}$, where $\sigma_0$ is the dark (background) conductivity, $A$ a conductivity amplitude weighing factor, and $\tau$ a characteristic relaxation time. Using $(\tau) = k_B T \ln(\tau \nu)$[36,35] we extracted the activation energy $E$ for each decay process associated with the relaxation times $\tau$ determined from these fits. Here $k_B$ is the Boltzmann constant, $T$ denotes the temperature (300 K), and $\nu$ is the attempt frequency which we take to be a typical phonon frequency of $\nu = 5 \times 10^{13}$ Hz. The results of these fits are summarized in **Table 2**. While the STO sample annealed without SrO (Figure 3(a)) could be satisfactorily fitted using a double exponential function, all the data in Figure 3(b) and (c) need triple exponential functions for acceptable fits. The activation energies calculated from the obtained relaxation times lie between 0.82 and 1.11 eV, and are plotted as data points in **Figure 4**, with the corresponding relative amplitudes (right y-axis) given by the weighing factors $A_i$ of the individual exponential components. Clearly, these defect states are located deep in the 3.3 eV gap of STO. Comparing these activation energies to recently reported calculated energy levels within the STO band gap for the defects that could be expected from our annealing conditions,[18,21,28,32] reveals that the extracted activation energies are inconsistent with those of known defect levels (see supplementary information).

Our results therefore suggest that the presence of isolated, independent defects is unable to explain the measured decay of the photoconductivity but that interacting defect complexes should be considered instead.[20,23] In the case of a perovskite oxide such as STO, the number of possible defect complexes can be significant.[21] This is compounded in the present study by the likely presence of Si interstitials or $Si_{Ti}$ and $Si_{Sr}$ antisite defects, which to our knowledge have not yet been considered theoretically. In this context, it is well-known that relaxations in many disordered systems can be empirically described using a stretched exponential or Kohlrausch function, i.e. $(t) = \sigma_0 e^{-\left(\frac{t}{\tau}\right)^\beta}$, $0 < \beta < 1$.[38,39] While it is commonly assumed that a stretched exponential decay arises because there is a distribution of independent relaxation processes with associated activation energies, it was shown[39] that relaxation in interacting disordered materials with hierarchically constrained (i.e. cascading) dynamics also gives rise to stretched exponential decay, where the stretch exponent $\beta$ has been suggested to be the inverse of the number of interacting or cooperating processes that must occur to lead to a relaxation event.[40] By fitting our data with a stretched exponential function (included in Figure 3), even with the significantly reduced number of parameters, fits were obtained with a similar quality as compared to double and triple exponential functions. Table 2 lists the resulting stretch exponents, $\beta$, and relaxation times, $\tau$. The values for $\beta$ are relatively small (<0.55), consistent with a large deviation from a single exponential Debye decay. It is worth noting that for sample A, the double (independent) exponential fit suggests only two processes are active in the decay, which should have resulted in the largest value of $\beta$, i.e. the system is closest to a description with a single exponential decay ($\beta$=1). Instead, the stretched exponential fit results in an exceedingly small $\beta$=0.25. If we take $\beta^{-1}$ to be the number of cooperating or correlated processes[40,41] associated with specific defect structures, then the observed relaxation of the persistent photoconductivity for the defect configurations analyzed here should involve 2 to 4 defect states, see Table 2. The discrepancy between the number of independent exponentials for each individual dataset and the value of $\beta^{-1}$ needed to fit the data (Table 2) implies that the stretched exponential decays cannot be explained with a distribution of independent relaxation processes, each with its own relaxation time.

Finally, we have extracted effective activation energy distribution from these stretched exponential decays by calculating the inverse Laplace transform of the decaying photoconductivity.[42,43,44] These distributions are plotted as continuous curves in Figure 4 for the decays presented in Figure 3, normalized to have the same total area under each curve for better visibility. Comparing these peaked activation energy distributions with the monotonic increase in the relative amplitudes obtained from the multiple (independent) exponential fits (datapoints in Figure 4) shows that, again, there is a discrepancy between the independent defect model, and the correlated defect model and the data: the data points do not correspond to the envelope of the (continuous) distributions, and the activation energies obtained from the multiple exponential fits (datapoints in Figure 4) with the largest weight are significantly removed (> 100 meV) from the maxima in the distribution curves from the stretched exponential fits.

This leads us to the important conclusion that the observed relaxation of the photoconductivity in the STO system cannot be understood by a sum of independent individual processes each with their own activation energy. Instead, the relaxation proceeds through cooperative or hierarchical[24,39] events that correlate the relaxation processes occurring at different defects. Our main observations intuitively fit this picture: (*i*) the changing nature (characteristic relaxation time) of the trapping states with increased defect concentration (comparing sample C and $B_{vac}$, see Table 1 and 2) implies that the increase of the concentration of individual defects creates new complexes made up of different sets of interacting defects. The non-trivial way in which the associated defect states depend on the detailed structure and interaction of such complexes[23] naturally explains the changing characteristic relaxation times upon increasing the defect concentration by quenching the sample from the annealing temperature. (*ii*) The changing relaxation times upon decreasing only the oxygen vacancy concentration (comparing both samples B) corroborates this explanation; eliminating oxygen vacancies from a defect complex does not simply remove an independent defect level associated with an independent oxygen vacancy from the extracted activation energies (Figure 4). Instead, it alters or shifts the effective activation energy distribution, i.e. it changes the character of the trapping state complexes. (*iii*) The fact that the effective activation energies extracted from fits based on independent Debye decays do not correspond to deep trap states of the most likely point defects, nor to the distribution of activation energies extracted from the stretched exponential fit, indicates that the effective activation energies governing the dynamic decay present in the samples are indeed non-trivially affected by the interactions between the defects. We conjecture that while Coulomb interactions between charged defects may dominate these interactions in many materials, for the case of STO these interactions are likely mediated through lattice deformations as the high dielectric constant of STO screens the Coulomb interaction between charged defects forming a complex. Indeed, the STO lattice does significantly respond to the presence of defects as evidenced by the presence of polarons near defects.[18,28]

**Conclusions**

In our experiments we have shown that photo-activated electron transport in STO single crystals with low defect concentrations cannot be described by a superposition of properties due to independent point defects as current models suggest. Instead, defect complexes that interact through dynamic correlations determine the effective activation energies for electronic transport. Our experiments therefore reveal that in order to understand how defects affect the electronic properties of STO, its interfaces, and complex oxides and heterostructures in general, it is necessary to consider models beyond those based on independent defects. Our finding bears important implications for defect engineering of complex oxides to obtain desired properties in heterogeneous catalysis and low dimensional electron systems, as well as for gaining an understanding of superconductivity or colossal magnetoresistance in these materials where similar interactions lead to emergent properties. Moreover, the approach demonstrated in this work makes it possible to gain spectroscopic insight into the defect complexes that dominate electronic properties in complex oxides through basic transport experiments.

**Experimental Methods**

5 mm x 5 mm x 0.5 mm $SrTiO_3$ single crystal samples were used in this study. To produce different defect complexes, we placed these samples in different environments during a 1 h 1200 °C anneal in a tube furnace with a $Al_2O_3$ tube. Apart from creating oxygen vacancies by annealing in vacuum, Si ampoules and $Al_2O_3$ boats were used to investigate the influence of possible Si-induced defects. Finally, SrO powder was added to prevent STO decomposition during high temperature annealing, and create Ti vacancies.[17] Specifically, three samples were placed in sealed $SiO_2$ ampoules evacuated to < 1 mTorr; two of these ampoules were also filled with ~2 grams of SrO powder. The fourth sample was placed in an $Al_2O_3$ boat with ~2 grams SrO powder within the tube furnace's sample space that was evacuated to < 1 mTorr during annealing. All samples were heated from 20 °C to 1200 °C at 10 °C $min^{-1}$ and all samples were cooled at 10 °C $min^{-1}$ with the exception of one Si ampoule-SrO powder sample which was quench cooled to 20 °C by directly removing from 1200 °C furnace.

4-probe electrical contacts were fabricated using indium solder with a separation of 3-4 mm between contacts. Photoexcitation was implemented by locating two 405 nm (3.06 eV) ultraviolet LEDs with a luminous intensity of 160 mcd 10 mm above the sample surface. The attenuation of this below band gap light is small, and the photoexcitation thus takes place throughout the bulk of the sample. This device was placed in a vacuum chamber with a base pressure of $10^{-8}$ Torr. Photoconductivity was measured at room temperature with a constant current mode of 10 μA. X-ray Photoelectron spectra were recorded in ultra-high vacuum using monochromatized Al K$α$ x-rays after a mild degassing at 200 °C for 30 minutes. A Shirley background was subtracted from the spectra.

**Acknowledgements**

PCS thanks T. Egami for insightful discussions. Research sponsored by the Laboratory Directed Research and Development Program of Oak Ridge National Laboratory, managed by UT-Battelle, LLC, for the U. S. Department of Energy. We acknowledge partial support from the U.S. Department of Energy, Office of Science, Basic Energy Sciences, Materials Sciences and Engineering Division (MPM, AH, AW). Y.-Z.M. was supported by the U.S. Department of Energy, Office of Science, Basic Energy Sciences, Chemical Sciences, Geosciences, and Biosciences Division.

**Author contributions**

TZW fabricated samples with help from AFM, and carried out the transport experiments with help from ATW and with input from PCS on annealing treatments. CS performed the inverse Laplace transformations. MPM carried out the XPS experiments. AH did the XRD measurements. PCS analyzed the data and conceived the interpretation. PCS and TZW supervised the project and wrote the manuscript. YZM contributed to understanding optical properties of defect states. All authors contributed to discussions and commented on the manuscript.

The authors declare no competing financial interests.

[1] Von Klitzing, K. The quantized Hall effect. *Rev. Mod. Phys.* **58**, 519-531 (1986).

[2] Stormer, H.L. Nobel lecture: the fractional quantum Hall effect. *Rev. Mod. Phys.* **71**, 875-889 (1999).

[3] Lee, C.-H. et al. Exploiting dimensionality and defect mitigation to create tunable microwave dielectrics. *Nature* **502**, 532-536 (2013).

[4] Waser, R., Dittmann, R., Staikov, G., Szot, K. Redox-based resistive switching memories – nanoionic mechanisms, prospects, and challenges. *Adv. Mat.* **21**, 2632-2663 (2009).

[5] Mikheev, E., Hwang, J., Kajdos, A.P., Hauser, A.J., Stemmer, S. Tailoring resistive switching in $Pt/SrTiO_3$ junctions by stoichiometry control. *Sci. Rep.* **5**, 11079 (2015).

[6] W.D. Rice, et al., Persistent optically induced magnetism in oxygen-deficient strontium titanate. *Nature Mater.* **13**, 481-487 (2013).

[7] Schooley, J.F. et al., Dependence of the superconducting transition temperature on carrier concentration in semiconducting $SrTiO_3$. *Phys. Rev. Lett.* **14**, 305-307 (1965).

[8] Jang, H.W., et al. Ferroelectricity in strain-free $SrTiO_3$ thin films. *Phys. Rev. Lett.* **104**, 197601 (2010).

[9] Chambers, S.A. Understanding the mechanism of conductivity at the $LaAlO_3/SrTiO_3$(001) interface. *Surf. Sci.* **605**, 1133-1140 (2011).

[10] Warusawithana, M.P. et al. $LaAlO_3$ stoichiometry is key to electron liquid formation at $LaAlO_3/SrTiO_3$ interfaces. *Nat. Commun.* **4**, 2351 (2013).

[11] Herranz, G. et al. High mobility in $LaAlO_3/SrTiO_3$ heterostructures: origin, dimensionality, and perspectives. *Phys. Rev. Lett.* **98**, 216803 (2007).

[12] Liu, Z.Q. et al., Origin of the two-dimensional electron gas at LaAlO$_3$/SrTiO$_3$ interfaces: the role of oxygen vacancies and electronic reconstruction. *Phys. Rev. X* **3**, 021010 (2013).

[13] Mueller, D.N., Machala, M.L., Bluhm, H., Chueh, H.C. Redox activity of surface oxygen anions in oxygen-deficient perovskite oxides during electrochemical reactions. *Nat. Commun*. **6**, 6097 (2015).

[14] Tuller, H.L., Bishop, S.R. Point defects in oxides: tailoring materials through defect engineering. *Annu. Rev. Mater. Res.* **41**, 369-398 (2011).

[15] Lang, D.V., Logan, R.A. Large-lattice-relaxation model for persistent photoconductivity in compound semiconductors. *Phys. Rev. Lett.* **39**, 635-639 (1977).

[16] Nathan, M.I. Persistent photoconductivity in AlGaAs/GaAs modulation doped layers and field effect transistors: a review. *Sol. Stat. Electr.* **29**, 167-172 (1986).

[17] Tarun, M.C., Selim, F.A., McCluskey, M.D. Persistent photoconductivity in strontium titanate. *Phys. Rev. Lett*. **111**, 187403 (2013).

[18] Janotti, A., Varley, J.B., Choi, M., Van de Walle, C.G. Vacancies and small polarons in SrTiO$_3$. *Phys. Rev. B* **90**, 085202 (2014).

[19] Ertekin, E. et al. Interplay between intrinsic defects, doping, and free carrier concentration in SrTiO$_3$ thin films. *Phys. Rev. B* **85**, 195460 (2012).

[20] Neugebauer, J. Van de Walle C.G. Gallium vacancies and the yellow luminescence in GaN. *Appl. Phys. Lett*. **69**, 503-505 (1996).

[21] Liu, B. et al., Composition dependent intrinsic defect structures in SrTiO$_3$. *Phys. Chem. Chem. Phys*. **16**, 15590-15596 (2014).

[22] Kim, Y., Disa, A.S., Babakol, T.E., Fang, X., Brock, J.D. Strain and oxygen vacancy ordering in SrTiO$_3$: Diffuse x-ray scattering studies. *Phys. Rev. B* **92**, 064105 (2015).

[23] Raebiger, H. Theory of defect complexes in insulators. *Phys. Rev. B* **82**, 073104 (2010).

[24] Fan, Y. Iwashita, T., Egami, T. Crossover from localized to cascade relaxations in metallic glasses. *Phys. Rev. Lett*. **115**, 045501 (2015).

[25] Ohtomo, A. Hwang, H.Y. A high-mobility electron gas at the LaAlO$_3$/SrTiO$_3$ heterointerface. *Nature* **427**, 423-426 (2004).

[26] Akhtar, M.J. Akhtar, Z.-U.-N., Jackson, R.A. Computer simulation studies of strontium titanate. *J. Am. Ceram. Soc.* **78**, 412-428 (1995).

[27] Moos, R., Menesklou, W., Härdtl, K.H. Hall mobility of undoped n-type conducting strontium titanate single crystals between 19 K and 1373 K. *Appl. Phys. A* **61**, 389-395 (1995).

[28] Hao, X. Wang, Z., Schmid, M., Diebold, U., Franchini, C. Coexistence of trapped and free excess electrons in SrTiO$_3$. *Phys. Rev. B* **91**, 085204 (2015).

[29] Lamoreaux, R.H., Hildenbrand, D.L. High-temperature vaporization behavior of oxides II. Oxides of Be, Mg, Ca, Sr, Ba, B, Al, Ga, In, Tl, Si, Ge, Sn, Pb, Zn, Cd and Hg. *J. Phys. Chem. Ref. Data* **16**, 419-443 (1987).

[30] Scofield, J.H. Hartree-Slater subshell photoionization cross-sections at 1254 and 1487 eV. *J. Electr. Spectrosc. Rel. Phenom.* **8**, 129-137 (1976).

[31] Hanzig, F. et al., Crystallization dynamics and interface stability of strontium titanate thin films on silicon. *J. Appl. Cryst.* **48**, 393-400 (2015).

[32] Tanaka, T., Matsunaga, K., Ikuhara, Y., Yamamoto, T. First-principles study on structures and energetics of intrinsic vacancies in SrTiO$_3$. *Phys. Rev. B* **68**, 205213 (2003).

[33] Santander-Syro, A.F.. et al. Two-dimensional electron gas with universal subbands at the surface of SrTiO$_3$. *Nature* **469**, 189-193 (2011).

[34] Meevasana, W. et al. Creation and control of a two-dimensional electron liquid at the bare SrTiO$_3$ surface. *Nat. Mater.* **10**, 114-118 (2011).

[35] McKeown Walker, S. et al. Control of a two-dimensional electron gas on $SrTiO_3$(111) by atomic oxygen. *Phys. Rev. Lett.* **113**, 177601 (2014).

[36] Studenikin, S.A., Golego, N., Cocivera, M. Optical and electrical properties of undoped ZnO films grown by spray pyrolysis of zinc nitrate solution. *J. Appl. Phys.* **83**, 2104-2111 (1998).

[37] Lin, J.Y., Dissanayake, A., Brown, G., Jiang, H.X. Relaxation of persistent photoconductivity in $Al_{0.3}Ga_{0.7}As$. *Phys. Rev. B* **42**, 5855-5858 (1990).

[38] Kohlrausch, R. Theorie des elektrischen Rückstandes in der Leidner Flasche II. *Pogg. Ann. Phys. Chem.* **91**, 179-214 (1854).

[39] Palmer, R.D., Stein, D.L., Abrahams, E., Anderson, P.W. Models of hierarchically constrained dynamics for glassy relaxation. *Phys. Rev. Lett.* **53**, 958-961 (1984).

[40] Rault, J. Remarks on the Kohlrausch exponent and the Vogel-Fulcher-Tamann law in glass-forming materials. *J. Non-Cryst. Sol.* **260**, 164-166 (1999).

[41] Rault, J. Glass: Kohlrausch exponent, fragility, anharmonicity. *Eur. Phys. J. E* **35**, 26 (2012).

[42] Berberan-Santos, M.N. A luminescence decay function encompassing the stretched exponential and the compressed hyperbola. *Chem. Phys. Lett.* **460**, 146-150 (2008).

[43] Berberan-Santos, M.N., Bodunov, E.N., Valeur, B. Mathematical functions for the analysis of luminescence decays with underlying distributions 1. Kohlrausch decay function (stretched exponential). *Chem. Phys.* **315**, 171-182 (2005).

[44] Luo, J. et al., Transient photoresponse in amorphous In-Ga-Zn-O thin films under stretched exponential analysis. *J. Appl.Phys.* **113**, 153709 (2013).

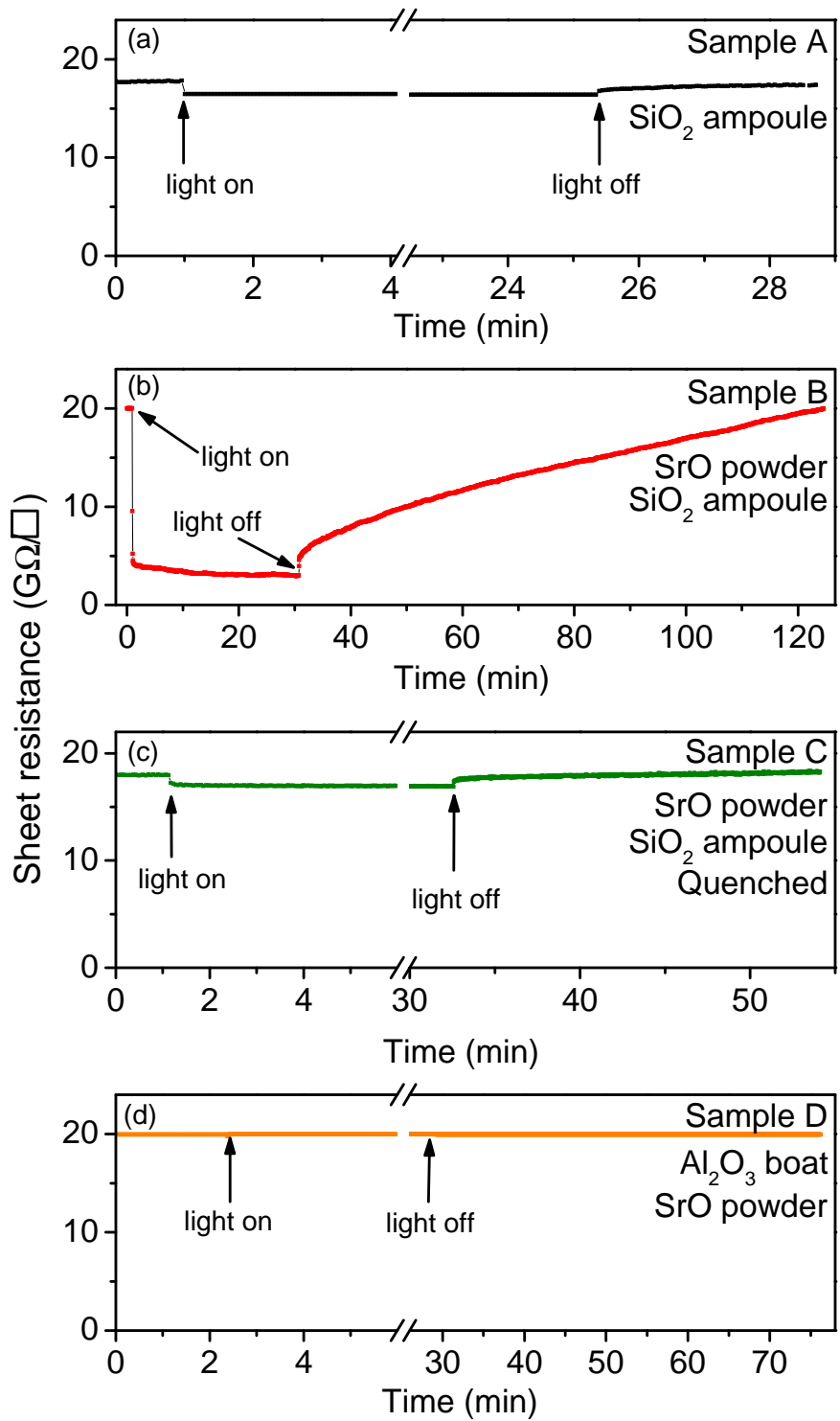

**Figure 1.** Changes in the electrical sheet resistance in SrTiO$_3$ single crystals as a function of time during and after illumination. In (a), (b), and (c) photoconductivity is observed, while it is absent in (d). The decay times vary for the different samples, depending on the annealing conditions, see also Table 1.

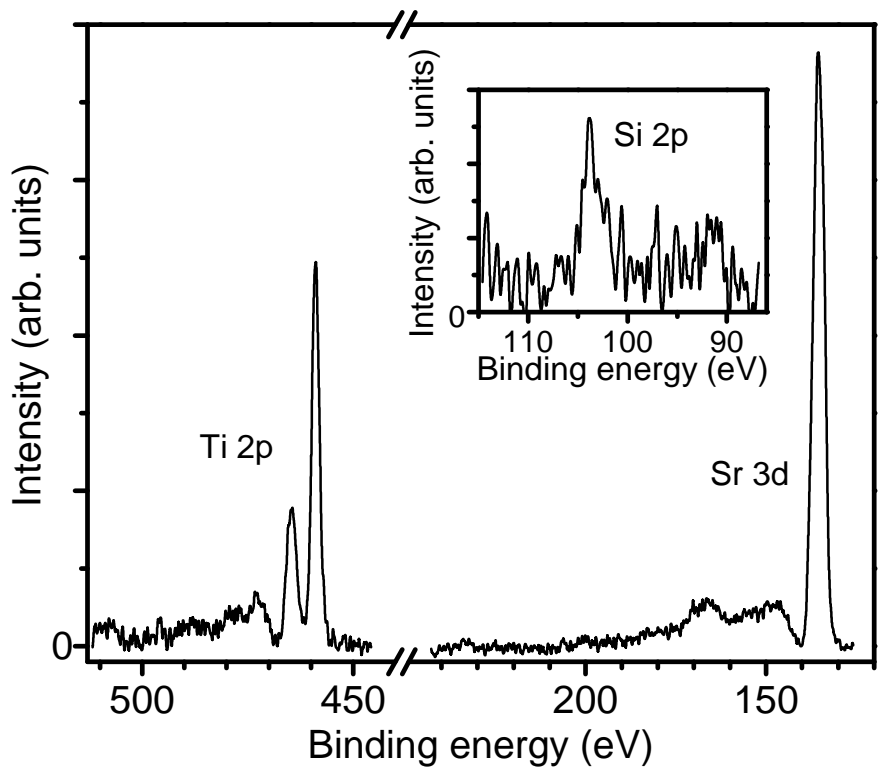

**Figure 2**. Photoelectron spectroscopy data for the Ti 2*p*, Sr 3*d*, and Si 2*p* (inset) core levels measured on sample B. From the integrated intensities of the Ti 2*p* and Sr 3*d* peaks, a Sr/Ti ratio of 3.1 is inferred, suggesting excess Sr is present on the surface. The data in the inset reveal the presence of Si on the surface of the sample.

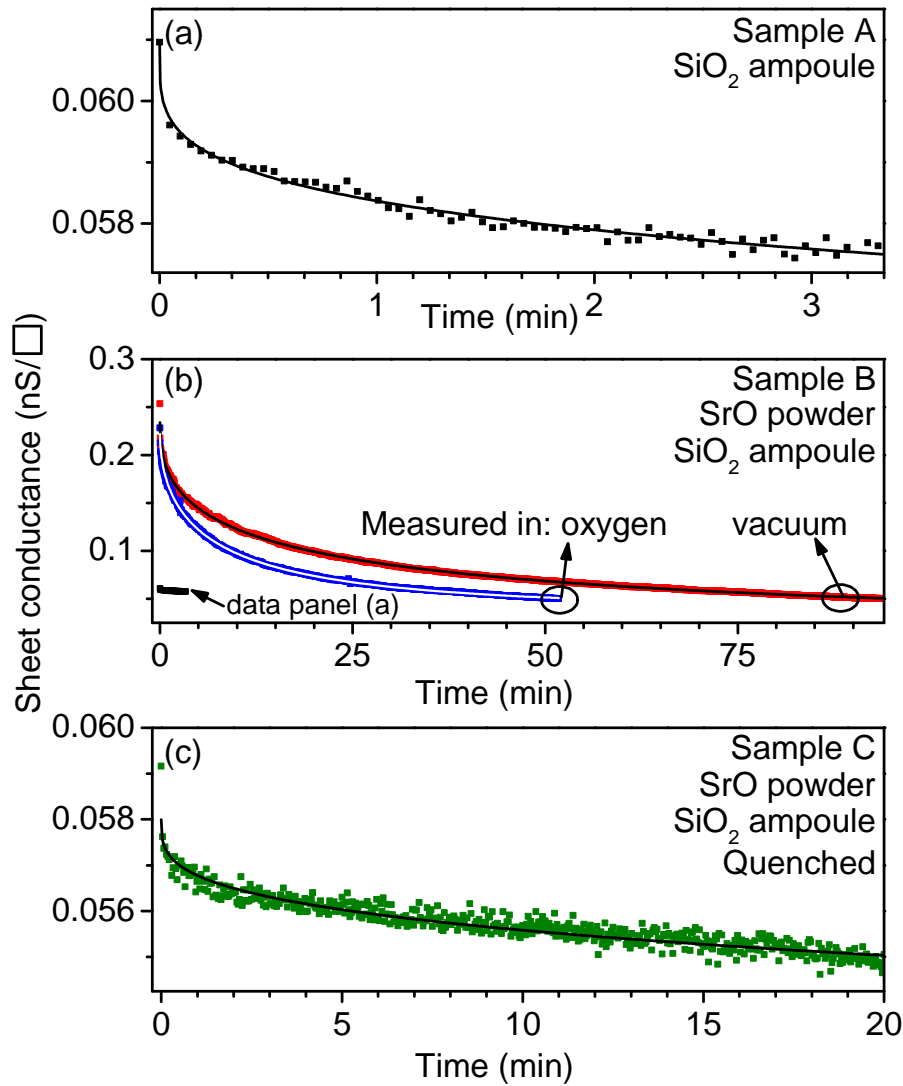

**Figure 3**. Decaying sheet photoconductance after the illumination is switched off. (a)-(c) present data for samples A-C, measured in vacuum. In (b), a measurement in an oxygen atmosphere is shown for comparison. Significantly different relaxation times are apparent, originating from different defects present in the samples. The data are fitted to stretched exponential functions plotted as continuous lines as discussed in the text.

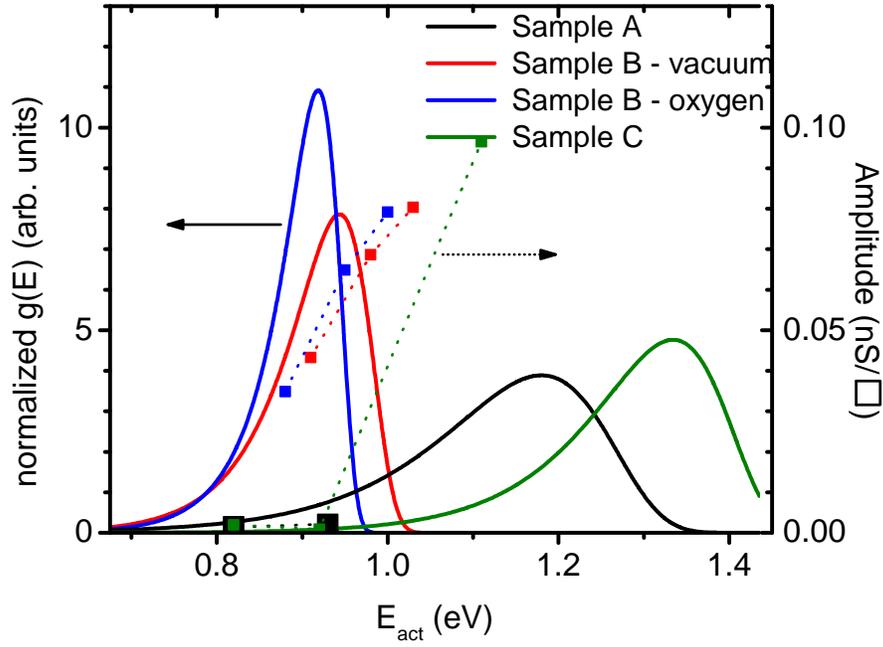

**Figure 4**. Activation energies for the processes dominating the decay in photoconductivity for samples A-C extracted using the two methods discussed in the text. The data points, connected by dotted lines, are calculated directly from the relaxation times obtained from fitting the decay curves with multiple exponential functions. The continuous curves are obtained through an inverse Laplace transformation of the stretched exponential decay fits.

**Table 1.** The annealing conditions, cooling rates, and the resulting expected defect types in the STO samples.

| Table 1 Annealing conditions | | | |
|---|---|---|---|
| Sample | Annealing environment | Cooling rate | Expected defects |
| A | SiO$_2$ Ampoule | 10K/min | $V_O$, Si imp. |
| B | SiO$_2$ Ampoule - SrO powder | 10K/min | $V_O$, Sr$_{Ti}$, $V_{Ti}$, Si imp. |
| C | SiO$_2$ Ampoule - SrO powder | Quenched | $V_O$, Sr$_{Ti}$, $V_{Ti}$, Si imp. |
| D | Al$_2$O$_3$ boat - SrO powder | 10K/min | $V_O$ |

**Table 2**. Fitting parameters from (stretched) exponential fits to the decay in photoconductivity of STO samples with low concentrations of intentional defects.

| Table 2 Decay parameters obtained from exponential fits | | | | | | | |
|---|---|---|---|---|---|---|---|
| Sample | Multiple exponential fits | | | | Stretched exponential fits | | |
|  | $A_i$ [S] | $\tau_i$ [s] | $E_i$ [eV] | $\chi^2_{red}$ | $\tau$ [s] | $\beta$ | $\chi^2_{red}$ |
| A | 1.5E-12<br>2.1E-12 | 1.4<br>78.4 | 0.82<br>0.93 | 7E-27 | 1.54E7 | 0.25 | 9E-27 |
| B$_{vac}$ | 4.3E-11<br>6.9E-11<br>8.0E-11 | 43.2<br>583.9<br>4074.2 | 0.91<br>0.98<br>1.03 | 9E-25 | 1.21E3 | 0.44 | 7E-25 |
| B$_{oxy}$ | 3.5E-11<br>6.5E-11<br>7.9E-11 | 11.0<br>211.0<br>1186.2 | 0.88<br>0.95<br>1.00 | 4E-25 | 4.62E3 | 0.54 | 4E-25 |
| C | 1.8E-12<br>9.1E-12<br>9.7E-11 | 1.3<br>60.7<br>76771.9 | 0.82<br>0.92<br>1.11 | 2E-26 | 7.3E9 | 0.30 | 3E-26 |

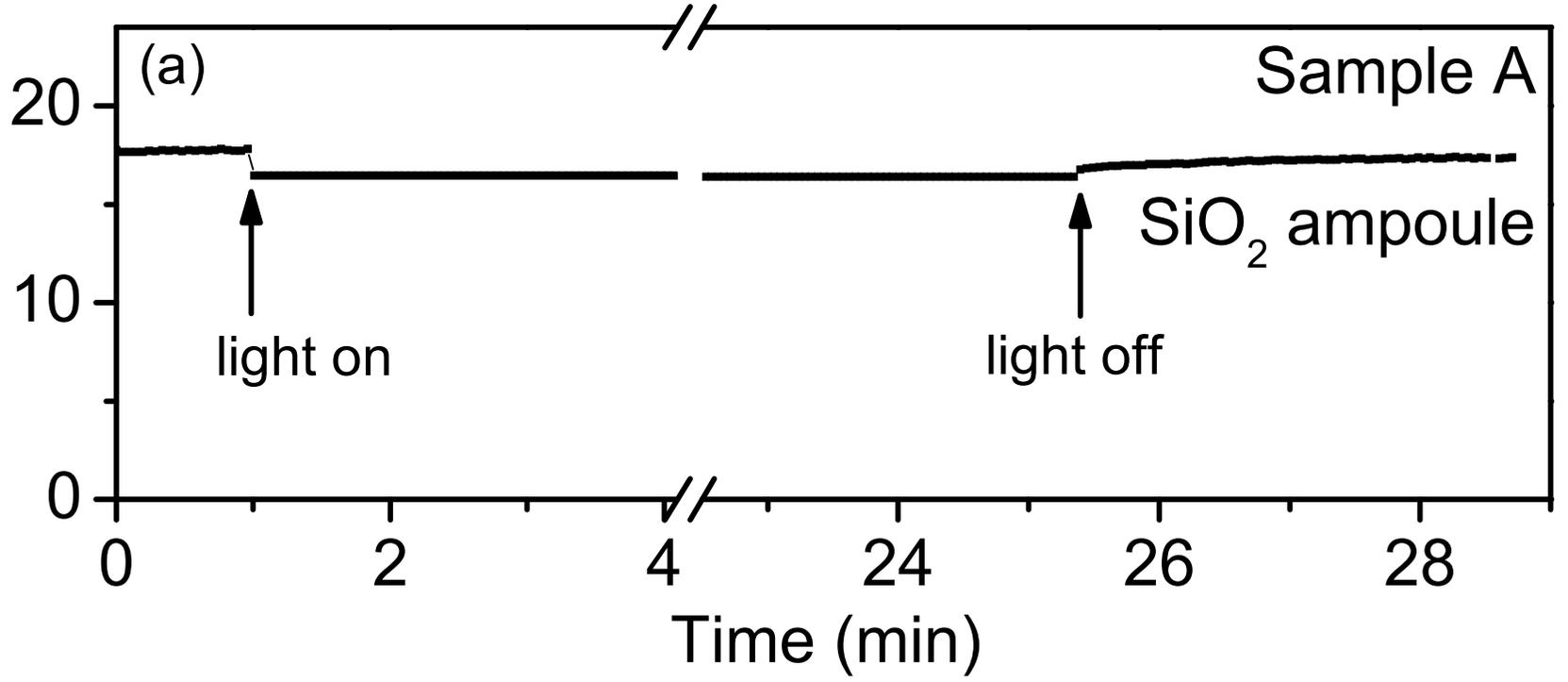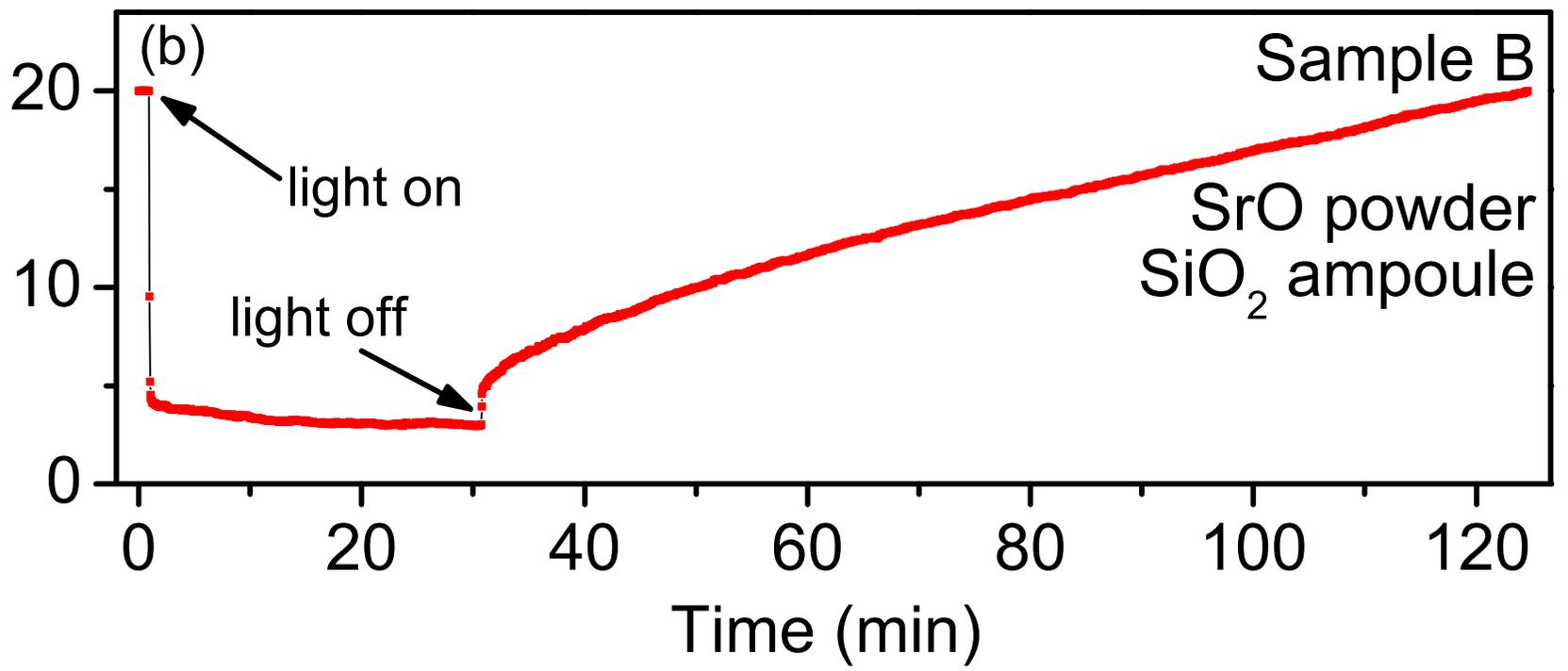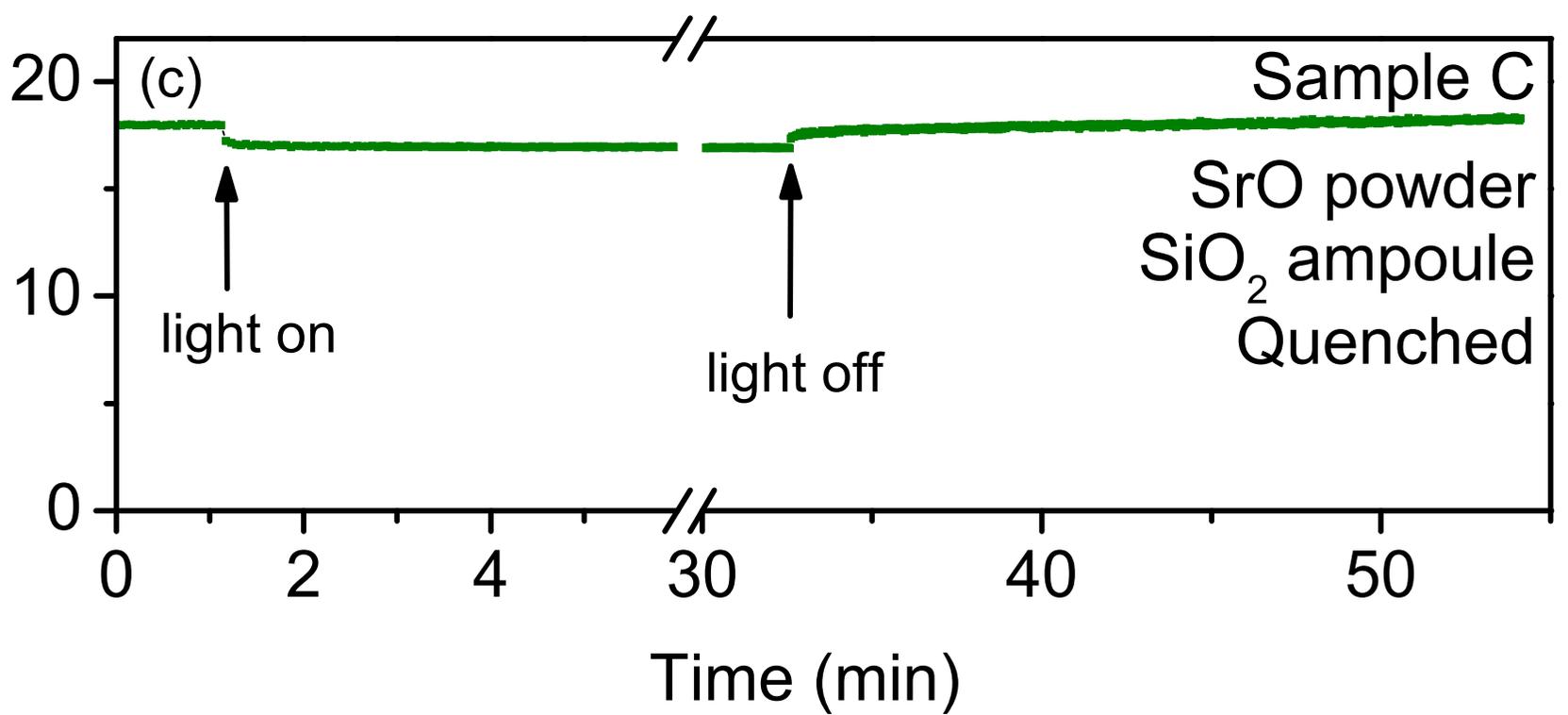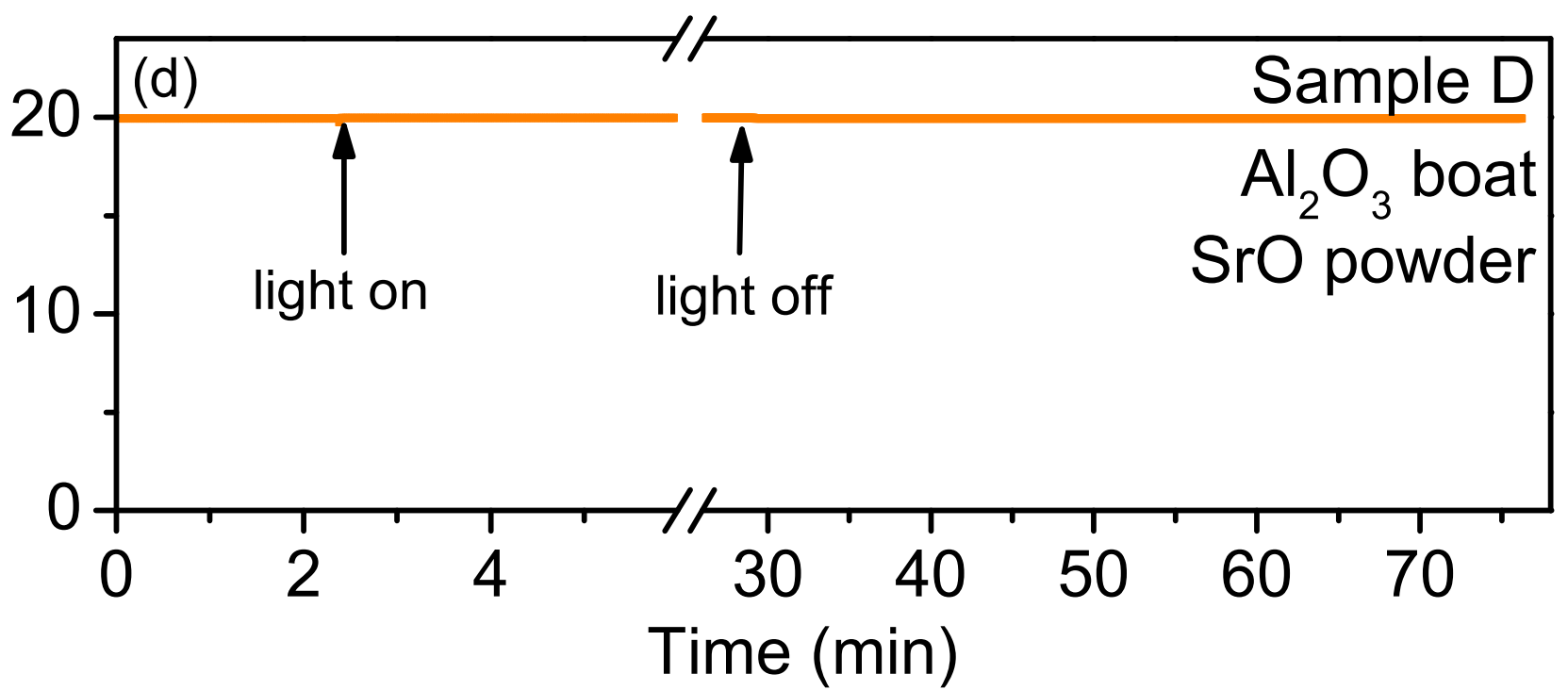

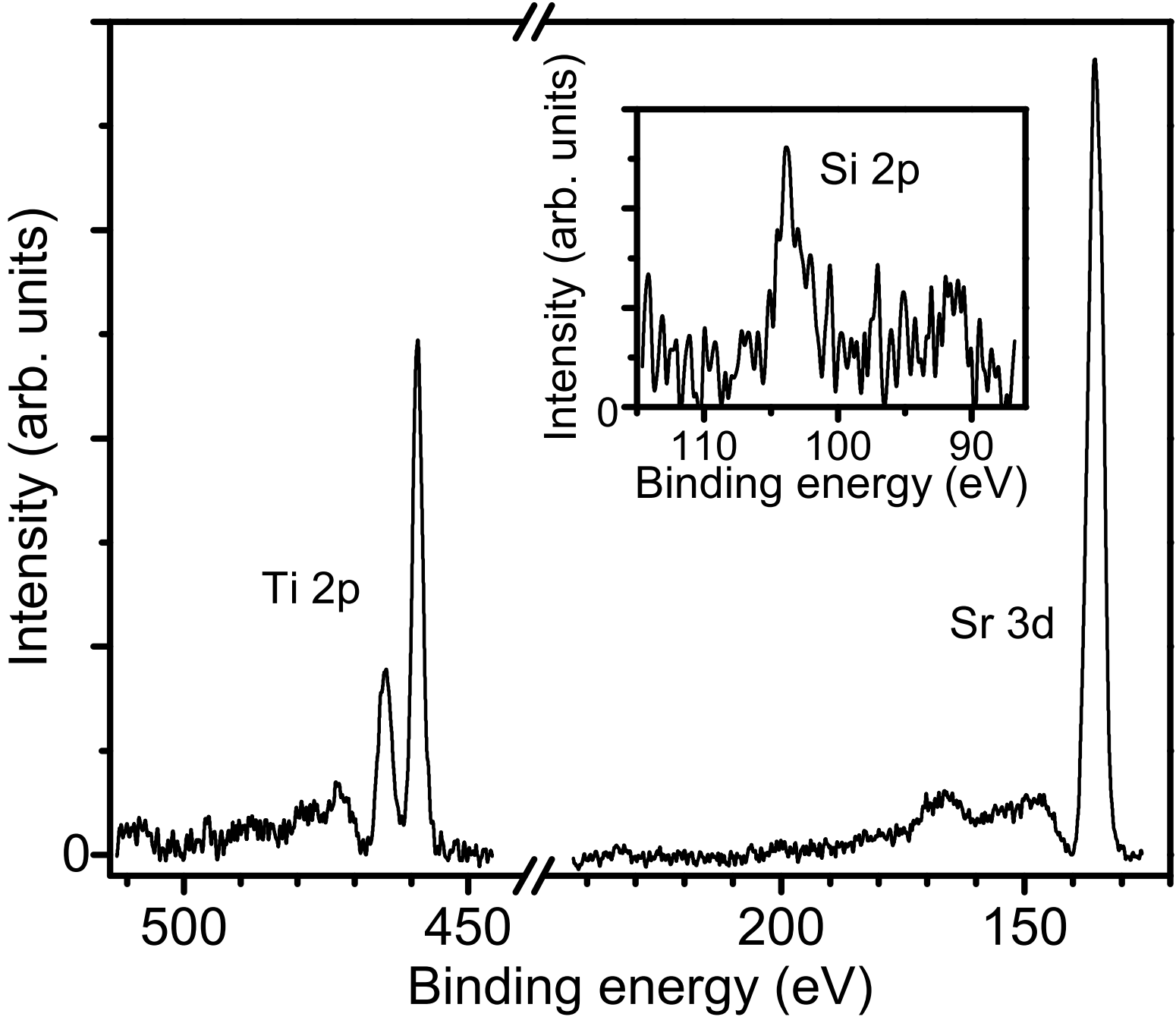

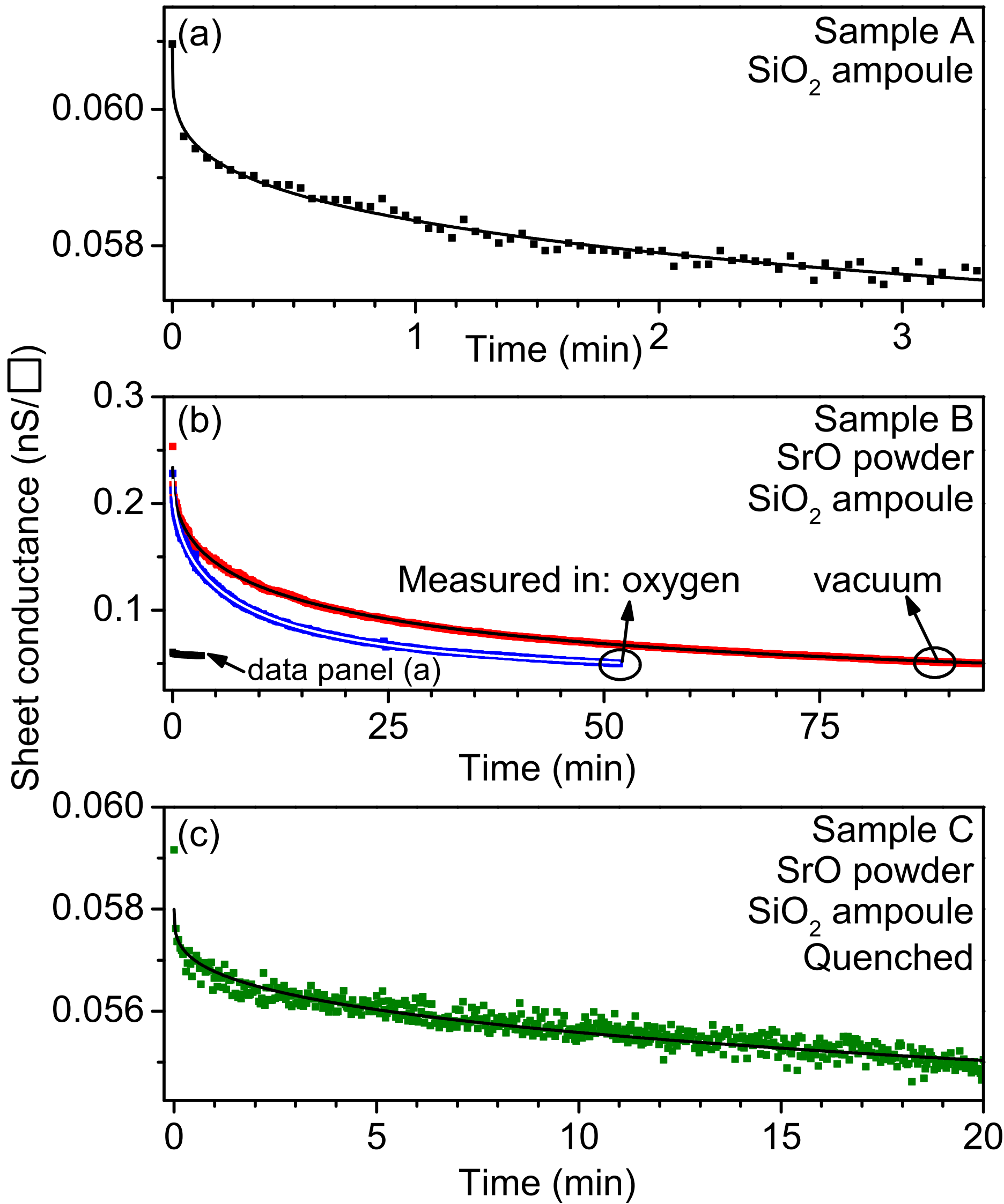

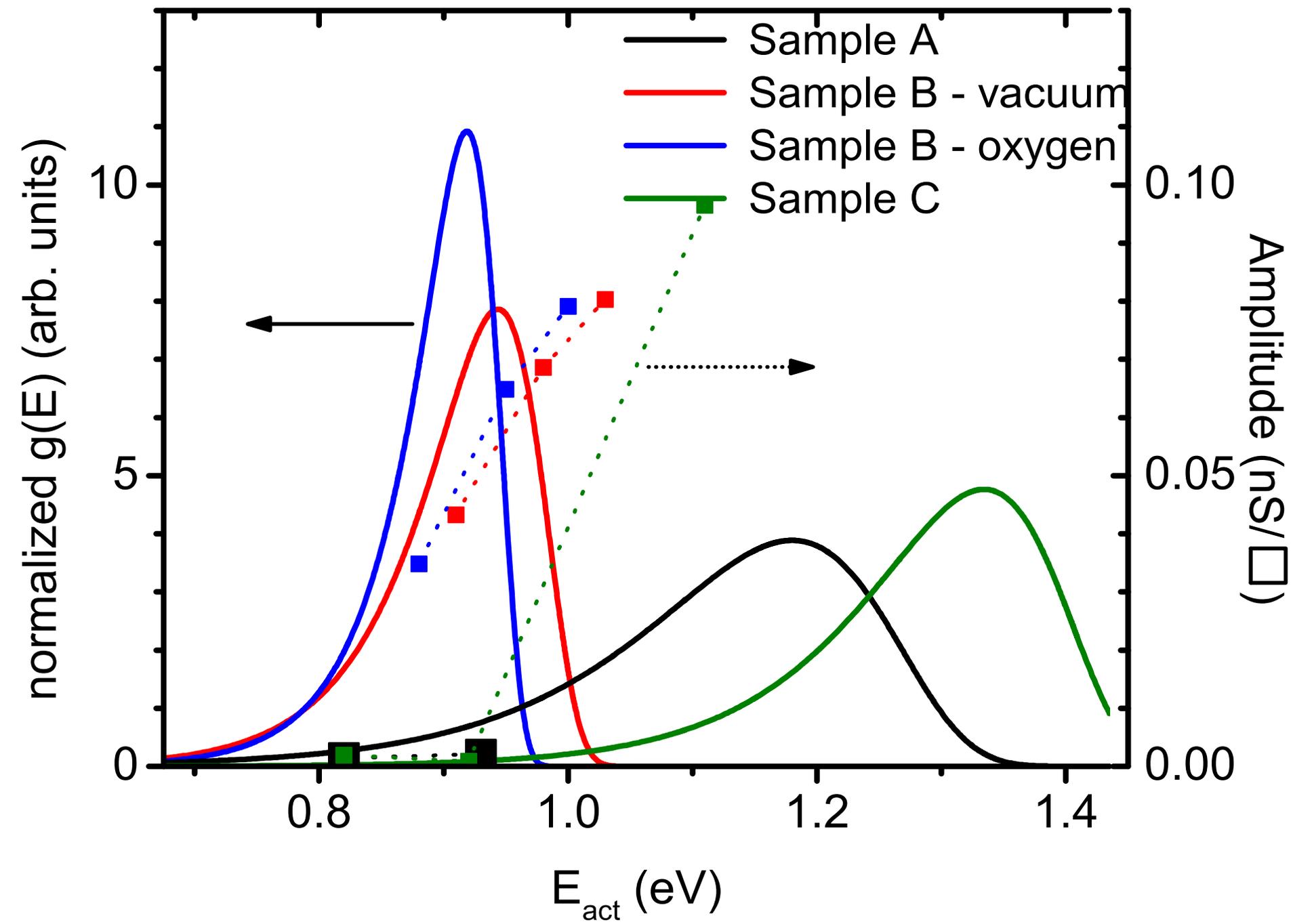

**Supplementary information for "Dynamic defect correlations dominate activated electronic transport in SrTiO$_3$**

*Paul C. Snijders,* Cengiz Şen, Michael P. McConnell, Ying-Zhong Ma, Andrew F. May, Andreas Herklotz, Anthony T. Wong*, and *T. Zac Ward**

**X-ray diffraction**

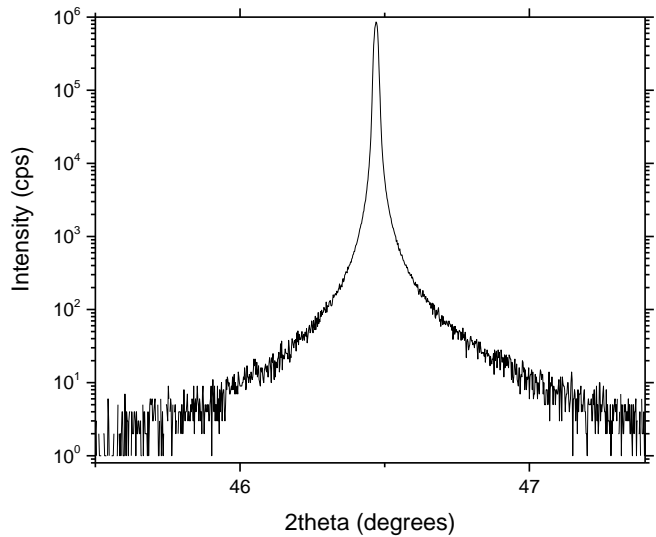

θ-2θ scan of sample C. Oxygen vacancies as well as cation off-stoichiometry[S2] will increase the STO lattice constant above its bulk. Both the out of plane (c) lattice constant of 3.905 Å, i.e. equal to that of relaxed bulk STO, and the absence of shoulders in the (002) reflection show that no defect-induced unit cell volume increases are present within the detection limit of X-ray diffraction, even for this quenched sample C that presumably has the highest defect concentration.

**Comparison of experimental relaxation data with results from Density Functional Theory (DFT) calculations.**

As discussed in the main text (Table 1), our annealing conditions are expected to create O and Ti vacancies, as well as $Sr_{Ti}$ antisite defects. In an attempt to identify the defects dominating the observed decay in photoconductivity, we compare our experimental data with the results of recent DFT calculations reported in the literature.[S2,S3,S4,S5] While we recognize that this comparison is limited by the small number of defect (complexes) that have been studied using state of the art DFT calculations, and potentially suffers from inaccuracies in DFT band gap determinations, DFT data are the only source of a direct relation between a specific defect type and the energy of its defect state within the STO band gap; experimental methods either average spectroscopic data over many different defects, or introduce the very same uncertainties by the need to be aided by theoretical (DFT) calculations to interpret atomically resolved spectroscopic data.

**Oxygen vacancies.**

Recent DFT calculations predict[S2] that oxygen vacancies are most stable when doubly ionized and forming a complex with a small polaron. This produces a localized, singly filled defect state located in the gap about 0.4 to 0.5 eV below the conduction band, and puts one electron in the conduction band. Clearly the energetic position of this level, as well as the conductivity that should arise from the induced conduction band carrier density is not observed in our measurements, and hence we believe a model based on independent oxygen vacancies does not describe our results. Another DFT study predicts[S4] that for very high carrier densities $\geq 10^{20}$ cm$^{-3}$, two electrons can be trapped near an oxygen vacancy by two small polarons. Below this critical carrier density, the system contains delocalized charge carriers or delocalized large polarons. As our samples were still visually transparent (except for the quench-cooled sample C) and highly resistive, we conclude that we did not create a defect density larger than this critical density. Our samples should therefore conduct reasonably well due to the polaron delocalization at low carrier densities, but instead they are highly resistive, implying that the defect configuration in our samples has to differ from those considered in Ref. [S4].

**Ti vacancies.**

While there are not many studies considering Ti vacancies, presumably because of their high formation energy,[S2,S5] the formation energy of Ti vacancies should decrease when SrO is present.[S3,S5] DFT calculations show that Ti vacancies are most stable as deep acceptors with a -4 charge state.[S2,S3] Upon photoexcitation with light just below the bandgap, an electron will be excited from this acceptor state into the conduction band. While recombination of the electron in the conduction band with the hole on the defect state does result in a photon emission peak at 1.2 eV because the state of $V_{Ti}^{-3}$ is located 1.2 eV below the conduction band,[S2] there is no energetic barrier for this capture process, which is necessary to explain the observed long term persistence of the photoconductivity - unless the hole on the Ti vacancy recombines with an electron from the valence band, but this corresponds to a transition energy of 2.1 eV, and leaves an excited electron in the conduction band. The latter recombination would not result in the observed decreasing persistent conductivity. As a result, we conclude that isolated Ti vacancies are not likely candidates to explain the observed persistent photoconductivity.

**Sr$_{Ti}$ antisite defects.**

Finally, a recent comprehensive study on defect complexes in STO[S3] considers the presence of excess SrO, which is comparable to our experimental conditions. It was shown that oxygen vacancies and Sr$_{Ti}^{-2}$ antisite defects are most prevalent under these conditions, but no energetics for in-gap states were provided. However, the generally lower formation energies of the considered defect complexes as compared to isolated point defects, points to the critical importance of such complexes as compared to individual defects in STO.

**Supplementary references**

[S1] Keeble, D.J., Wicklein, S., Jin, L., Jia, C.L., Egger, W., and Dittmann, R., Nonstoichiometry accommodation in SrTiO$_3$ thin films studied by positron annihilation and electron microscopy. *Phys. Rev. B* **87**, 195409 (2013).

[S2] Janotti, A., Varley, J.B., Choi, M., Van de Walle, C.G. Vacancies and small polarons in SrTiO$_3$. *Phys. Rev. B* **90**, 085202 (2014).

[S3] Liu, B., Cooper, V.R., Xu, H., Xiao, H., Zhang, Y., Weber, W.J. Composition dependent intrinsic defect structures in SrTiO$_3$. *Phys. Chem. Chem. Phys*. **16**, 15590-15596 (2014).

[S4] Hao, X. Wang, Z., Schmid, M., Diebold, U., Franchini, C. Coexistence of trapped and free excess electrons in SrTiO$_3$. *Phys. Rev. B* **91**, 085204 (2015).

[S5] Ohtomo, A. Hwang, H.Y. A high-mobility electron gas at the LaAlO$_3$/SrTiO$_3$ heterointerface. *Nature* **427**, 423-426 (2004).